# Formulating Initial and Boundary Effects for Maxwell Equations

*Xiao Jianhua*
*Natural Science Foundation Research Group, Shanghai Jiaotong University, Shanghai, P.R.C.*

**Abstract.** In classical treatment of Maxwell equations, the initial and boundary conditions are introduced by mathematical consideration rather than strictly using the Maxwell equations. As a result, the initial and boundary conditions are not logic consistent with the Maxwell equations. In fact, the boundary problem of classical Maxwell equations is not solvable for general cases, if the whole boundary fields are given. This problem is caused by implying the divergence and the curl of a vector have no relations during introducing the boundary condition. This research introduces new ideas and formulations to solve this problem. The research shows that based on Chen's S-R decomposition of a rank-two tensor, this logical un-consistency can be discarded and, as a consequence, the classical Maxwell equations are reformulated. According to the reformulated Maxwell equations, the initial condition and boundary condition have global effects. As an example, London equations for superconductors are deduced.

**Introduction**

For boundary problem, the classical Maxwell equations are not solvable mathematically, if the whole boundary fields are given. In general, when the normal components are given the tangent components should be determined by solution or the vise versa. From mathematic consideration, if the whole boundary fields are given, the Maxwell equations are over-determined and hence are not-solvable. This shows that there are some intrinsic logic un-consistency between the Maxwell equations and initial-boundary conditions[1].

In recent years, much attention has been devoted to the computation of electromagnetic fields in bounded singular domains, which has non-smooth or non-convex boundary. This kind of geometrical singularities has been studied by Jr. Patrick Ciarlet [2]. Starting from boundary conditions for the Maxwell equations, it is not obvious to deduce boundary conditions. Viewing this, some approximate models have been developed[3].

More mathematically, semi-linear Maxwell equations[4] and non-linear Maxwell equations[5] are developed to treat the complicated field problem. However, in semi-linear and non-linear Maxwell equations, for the gauge potential related to the magnetic field, it is not possible to choose the Coulumb gauge to avoid indefiniteness.

The most striking treatment is to introduce magnetic current and magnetic charge densities to make the generalized Maxwell equations[6]. Although this method has not been supported by physical experiments, its logic value is significant.

For initial and boundary value problem, the singularities problem can be discarded by Chen's S-R additive decomposition theory of a rank-two tensor. Based on this decomposition, the divergence of a vector and its curl are definitely related [7-8]. That is to say, even for the static electromagnetic field, the divergence of a magnetic field may be not zero and the curl of an electric field may be not zero for the case that the static field is changed in space sharply. This physical topic is addressed in modern physical theory by introducing a curved space or, more abstract, in four-dimensional space-time. For most engineering cases, such beautiful theories are hardly used. Hence, to study the intrinsic feature of electromagnetic field, the classical Maxwell equations should be reformulated with new understanding mentioned above.

This paper will introduce Chen's S-R decomposition equation and the reformed Maxwell equations firstly. After that, related equations will be deduced and discussed with some detail. Then, the paper goes to establish the relationship equation between the field within material and the outside field. Finally, the geometry design principle for superconductor and super-insulator is mentioned as an application example of the reformed Maxwell equations.

## 1. Extending Chen's S-R Additive Decomposition Theorem to any rank-two tensor

For finite deformation, Chen Zhida has shown that a transformation $F_j^i + \delta_j^i$ can be decomposed into the addition of one symmetry tensor $S_j^i$ expressing stretching and one unit orthogonal tensor $R_j^i$ expressing local rotation [7-8]. For our purpose, Chen's S-R decomposition theorem can be expressed as:

$$F_j^i + \delta_j^i = S_j^i + R_j^i \tag{1}$$

Where, $\delta_j^i$ is Kronecker delta:

$$S_j^i = \frac{1}{2}(F_j^i + F_j^{iT}) - (1 - \cos\Theta)L_k^i L_j^k \tag{2}$$

$$R_j^i = \delta_j^i + \sin\Theta \cdot L_j^i + (1 - \cos\Theta)L_k^i L_j^k \tag{3}$$

$$L_j^i = \frac{1}{2\sin\Theta}(F_j^i - F_j^{iT}) \tag{4}$$

$$\sin\Theta = \frac{1}{2}[(F_2^1 - F_1^2)^2 + (F_3^2 - F_2^3)^2 + (F_1^3 - F_3^1)^2]^{\frac{1}{2}} \tag{5}$$

In above expressions, the parameter $\Theta$ is called local average rotation angel and tensor $L_j^k$ defines the local average rotation axis direction. The upper index T represents transpose.

But, the definition of local rotation angel is too strong to be applicable for the field changed in space sharply, as it requires the condition of:

$$\frac{1}{2}[(F_2^1 - F_1^2)^2 + (F_3^2 - F_2^3)^2 + (F_1^3 - F_3^1)^2]^{\frac{1}{2}} \le 1 \tag{6}$$

This problem can be overcome by defining the local average rotation angel $\theta$ as the following:

$$(\cos\theta)^{-2} = 1 + \frac{1}{4}[(F_2^1 - F_1^2)^2 + (F_3^2 - F_2^3)^2 + (F_1^3 - F_3^1)^2] \tag{7}$$

Now, we give details to improve other equations.

Firstly, we noticed that:

$$\det[\frac{1}{2}(F_j^i - F_j^{iT}) + \delta_j^i] = (\cos\theta)^{-2} \tag{8}$$

Secondly, we find that the unit orthogonal rotation tensor defined by $\frac{\cos\theta}{2}(F_j^i - F_j^{iT})$ is:

$$R_j^i = \delta_j^i + \sin\theta \cdot L_j^i + (1 - \cos\theta)L_k^i L_j^k \tag{9}$$

Where, the local average rotation axis direction tensor $L_j^k$ is:

$$L_j^i = \frac{\cos\theta}{2\sin\theta}(F_j^i - F_j^{iT}) \tag{10}$$

Hence, by equations (9) and (10), we get:

$$\frac{1}{2}(F_j^i - F_j^{iT}) + \delta_j^i = (\cos\theta)^{-1}R_j^i - (\frac{1}{\cos\theta} - 1)(L_k^i L_j^k + \delta_j^i) \tag{11}$$

We finally get Chen's S-R additive decomposition theorem as:

$$F_j^i + \delta_j^i = S_j^i + (\cos\theta)^{-1}R_j^i \tag{12}$$

Where:

$$S_j^i = \frac{1}{2}(F_j^i + F_j^{iT}) - (\frac{1}{\cos\theta} - 1)(L_k^i L_j^k + \delta_j^i) \tag{13}$$

$$(\cos\theta)^{-1}R_j^i = \delta_j^i + \frac{1}{2}(F_j^i - F_j^{iT}) + (\frac{1}{\cos\theta} - 1)(L_k^i L_j^k + \delta_j^i) \tag{14}$$

Others are defined by equations (7), (9), and (10). The equation (14) can be rewritten as:

$$(\cos\theta)^{-1}R_j^i = \delta_j^i + \frac{\sin\theta}{\cos\theta}L_j^i + (\frac{1}{\cos\theta} - 1)(L_k^i L_j^k + \delta_j^i) \tag{15}$$

Hence, Chen Zhida's S-R decomposition is extended to any rank-two tensors.

## 2. Tensor Expression of Divergence and Curl of Electromagnetic Field

To reform the classical Maxwell equations, the tensor expression of divergence and curl of a vector must be established firstly. Note that here the space is supposed as orthogonal standard three-dimensional space rather than the curvature space.

### 2.1 Boundary Effects

For electromagnetic field, the field any point can be viewed as a function of the field at its neighbouring point Q (denoted by lower index Q). For a vector field $E^i$, this functional relationship can be expressed as:

$$E^i = \frac{\partial E^i}{\partial E_Q^j}\frac{\partial E_Q^j}{\partial x^k}dx^k + E_Q^i \tag{16}$$

That is to say, the electromagnetic fields are locally linear fields. Omitting higher orders infinitesimals, taking the differential of the above equations, one will get:

$$\frac{\partial E^i}{\partial x^k} = \frac{\partial E^i}{\partial E_Q^j}\frac{\partial E_Q^j}{\partial x^k} + \frac{\partial E_Q^i}{\partial x^k} = (\frac{\partial E^i}{\partial E_Q^j}+\delta_j^i)\frac{\partial E_Q^j}{\partial x^k} = (E_j^i+\delta_j^i)\frac{\partial E_Q^j}{\partial x^k} \tag{17}$$

Appling Chen's decomposition equations (7),(9),(10),(13), and (15) into above equation, one get:

$$\frac{\partial E^i}{\partial x^k} = (\frac{\partial E^i}{\partial E_Q^j}+\delta_j^i)\frac{\partial E_Q^j}{\partial x^k} = [S_j^i + (\cos\theta)^{-1}\cdot R_j^i]\frac{\partial E_Q^j}{\partial x^k} \tag{18}$$

Hence, we have:

$$dE^i = (E_j^i+\delta_j^i)dE_Q^j = [S_j^i + (\cos\theta)^{-1}R_j^i]dE_Q^j \tag{19}$$

Where, the related tensors are defined as:

$$S_j^i = \frac{1}{2}(E_j^i+E_j^{iT}) - (\frac{1}{\cos\theta}-1)(L_k^iL_j^k+\delta_j^i) \tag{20}$$

$$(\cos\theta)^{-1}R_j^i = \delta_j^i + \frac{\sin\theta}{\cos\theta}L_j^i + (\frac{1}{\cos\theta}-1)(L_k^iL_j^k+\delta_j^i) \tag{21}$$

$$L_j^i = \frac{\cos\theta}{2\sin\theta}(E_j^i-E_j^{iT}) \tag{22}$$

within them the local average rotation angle is given by:

$$(\cos\theta)^{-2} = 1 + \frac{1}{4}[(E_2^1-E_1^2)^2 + (E_3^2-E_2^3)^2 + (E_1^3-E_3^1)^2] \tag{23}$$

It shows that the local variation of a vector field can be decomposed as the additive of a local stretching and a local average rotation.

Based on these equations, the divergence and the curl the fields, taking electric field $E^i$ as example, can be expressed as:

$$\nabla\cdot\vec{E} = \frac{\partial E^i}{\partial x^i} = S_j^i\frac{\partial E_Q^j}{\partial x^i} + (\cos\theta)^{-1}\cdot R_j^i\frac{\partial E_Q^j}{\partial x^i} \tag{24}$$

The curl of the field can be expressed as:

$$(\nabla\times\vec{E})^i = e_{ijk}\frac{\partial E^j}{\partial x^k} = e_{ijk}S_l^j\frac{\partial E_Q^l}{\partial x^k} + (\cos\theta)^{-1}\cdot e_{ijk}R_l^j\frac{\partial E_Q^l}{\partial x^k} \tag{25}$$

For equation (24), if the field has zero divergence at point Q but its curl is not zero, that is:

$$\frac{\partial E_Q^i}{\partial x^j} = -\frac{\partial E_Q^j}{\partial x^i} \tag{26}$$

then, notice equations (20) and (21), using the symmetry feature of $S_j^i$, one will have:

$$\nabla\cdot\vec{E} = (\cos\theta)^{-1}R_j^i\frac{\partial E_Q^j}{\partial x^i} \tag{27}$$

It is clear that the divergence is not zero if the local average rotation angle is not zero at point Q. For magnetic field, it means that its divergence is not zero if the magnetic field variation has relative local rotation.

For equation (25), if the field has zero curl at point Q (its divergence is not zero), that is:

$$\frac{\partial E_Q^i}{\partial x^j} = \frac{\partial E_Q^j}{\partial x^i} \tag{28}$$

then, notice equations (20) and (21), one will get:

$$(\nabla \times \vec{E})^i = e_{ijk} \frac{\partial E^j}{\partial x^k} = (\cos\theta)^{-1} \cdot e_{ijk} R_l^j \frac{\partial E_Q^l}{\partial x^k} \tag{29}$$

It is clear that the curl is not zero if the local average rotation angle is not zero at point Q. For electric field, it means that its curl is not zero if the electrical field variation has relative local rotation. Such a kind of local rotation usually will appear at the points neighbouring boundary.

*2.2 Initial State Effects*

For alternating electromagnetic field, understanding the lower index 0 as indicating a reference time $t_0$, the similar equations are available also as one can replace the spatial differential with the time differential. That is:

$$E^i = \frac{\partial E^i}{\partial E_0^j} \frac{\partial E_0^j}{\partial t} dt + E_0^i \tag{30}$$

$$\frac{\partial E^i}{\partial t} = (\frac{\partial E^i}{\partial E_0^j} + \delta_j^i)\frac{\partial E_0^j}{\partial t} = (\tilde{E}_j^i + \delta_j^i)\frac{\partial E_0^j}{\partial t} = [\tilde{S}_j^i + (\cos\tilde{\theta})^{-1} \cdot \tilde{R}_j^i]\frac{\partial E_0^j}{\partial t} \tag{31}$$

Where, the related tensors are defined as:

$$\tilde{S}_j^i = \frac{1}{2}(\tilde{E}_j^i + \tilde{E}_j^{iT}) - (\frac{1}{\cos\tilde{\theta}} - 1)(\tilde{L}_k^i \tilde{L}_j^k + \delta_j^i) \tag{32}$$

$$(\cos\tilde{\theta})^{-1}\tilde{R}_j^i = \delta_j^i + \frac{\sin\tilde{\theta}}{\cos\tilde{\theta}}\tilde{L}_j^i + (\frac{1}{\cos\tilde{\theta}} - 1)(\tilde{L}_k^i \tilde{L}_j^k + \delta_j^i) \tag{33}$$

$$\tilde{L}_j^i = \frac{\cos\tilde{\theta}}{2\sin\tilde{\theta}}(\tilde{E}_j^i - \tilde{E}_j^{iT}) \tag{34}$$

within them the local average rotation angle is given by:

$$(\cos\tilde{\theta})^{-2} = 1 + \frac{1}{4}[(\tilde{E}_2^1 - \tilde{E}_1^2)^2 + (\tilde{E}_3^2 - \tilde{E}_2^3)^2 + (\tilde{E}_1^3 - \tilde{E}_3^1)^2] \tag{35}$$

Based on these equations, the divergence and the curl the fields, taking electric field $E^i$ as example, can be expressed as:

$$\nabla \cdot \frac{\partial \vec{E}}{\partial t} = \frac{\partial^2 E^i}{\partial t \partial x^i} = \tilde{S}_j^i \frac{\partial^2 E_0^j}{\partial t \partial x^i} + (\cos\tilde{\theta})^{-1} \cdot \tilde{R}_j^i \frac{\partial^2 E_0^j}{\partial t \partial x^i} \tag{36}$$

The curl of the field can be expressed as:

$$(\nabla \times \frac{\partial \vec{E}}{\partial t})^i = e_{ijk}\frac{\partial^2 E^j}{\partial t \partial x^k} = e_{ijk}\tilde{S}_l^j\frac{\partial^2 E_0^l}{\partial t \partial x^k} + (\cos\tilde{\theta})^{-1} \cdot e_{ijk}\tilde{R}_l^j\frac{\partial^2 E_0^l}{\partial t \partial x^k} \tag{37}$$

For equation (36), if the initial field has zero divergence at initial time, but its curl is not zero, that is:

$$\frac{\partial^2 E_0^i}{\partial t \partial x^j} = -\frac{\partial^2 E_0^j}{\partial t \partial x^i} \tag{38}$$

then, notice equations (32) and (33), using the symmetry feature of $\tilde{S}_j^i$, one will have:

$$\nabla \cdot \frac{\partial \vec{E}}{\partial t} = (\cos\tilde{\theta})^{-1}\tilde{R}_j^i\frac{\partial^2 E_0^j}{\partial t \partial x^i} \tag{39}$$

It is clear that the divergence is not zero if the local average rotation angle is not zero at initial time. For magnetic field, it means that its divergence is not zero if the magnetic field variation has relative local rotation.

For equation (37), if the field has zero curl at initial time (its divergence is not zero), that is:

$$\frac{\partial^2 E_0^i}{\partial t \partial x^j} = \frac{\partial^2 E_0^j}{\partial t \partial x^i} \tag{40}$$

then, notice equations (32) and (33), one will get:

$$(\nabla \times \frac{\partial \vec{E}}{\partial t})^i = e_{ijk}\frac{\partial^2 E^j}{\partial t \partial x^k} = (\cos\tilde{\theta})^{-1} \cdot e_{ijk}\tilde{R}_l^j\frac{\partial^2 E_0^l}{\partial t \partial x^k} \tag{41}$$

It is clear that the curl is not zero if the local average rotation angle is not zero at initial time. For electric field, it means that its curl is not zero if the electrical field variation has relative local rotation. Such a kind of local rotation usually will appear in the processes of initial stage.

Summering above results, it can be concluded that, for any field, if its variation has relative local rotation then its divergence and curl will be coupled together.

**3. Reformulated Maxwell Equations**

Based on above conclusion, the divergence of magnetic field cannot be always zero. This means that the classical Maxwell equations, which require the divergence of any magnetic field is always zero, are not logic consistent.

The classical Maxwell equations are:

$$\nabla \cdot \vec{B} = 0 \tag{42}$$

$$\nabla \times \vec{E} + \frac{\partial \vec{B}}{\partial t} = 0 \tag{43}$$

$$\nabla \cdot \vec{D} = \sigma \tag{44}$$

$$\nabla \times \vec{H} = \frac{\partial \vec{D}}{\partial t} + \vec{J} \tag{45}$$

The medium feature is defined by:

$$\vec{D} = \varepsilon \vec{E}, \quad \vec{B} = \mu \vec{B}, \quad \vec{J} = \gamma \vec{E} \tag{46}$$

According to the results of last section, the logic un-consistency problem of classical Maxwell equations can be discarded by introducing correction items and reforming the equations as:

$$\nabla \cdot \vec{B} = \sigma_B \tag{47}$$

$$\nabla \times \vec{E} + \frac{\partial \vec{B}}{\partial t} = \vec{J}_E \tag{48}$$

$$\nabla \cdot \vec{D} = \sigma + \sigma_D \tag{49}$$

$$\nabla \times \vec{H} = \frac{\partial \vec{D}}{\partial t} + \vec{J} + \vec{J}_H \tag{50}$$

where:

$$\sigma_B = (\cos \tilde{\theta}_B)^{-1} \tilde{R}^i_{Bj} \frac{\partial B^j_0}{\partial x^i} + (\cos \theta_Q)^{-1} R^i_{Bj} \frac{\partial B^j_Q}{\partial x^i} \tag{51}$$

$$J^i_E = (\cos \tilde{\theta}_E)^{-1} \cdot e_{ijk} \tilde{R}^j_{El} \frac{\partial E^l_0}{\partial x^k} + (\cos \theta_E)^{-1} \cdot e_{ijk} R^j_{El} \frac{\partial E^l_Q}{\partial x^k} \tag{52}$$

$$\sigma_D = (\cos \tilde{\theta}_D)^{-1} \tilde{R}^i_{Dj} \frac{\partial D^j_0}{\partial x^i} + (\cos \theta_D) R^i_{Dj} \frac{\partial D^j_Q}{\partial x^i} \tag{53}$$

$$J^i_H = (\cos \tilde{\theta}_H)^{-1} \cdot e_{ijk} \tilde{R}^j_{Hl} \frac{\partial H^l_0}{\partial x^k} + (\cos \theta_H) \cdot e_{ijk} R^j_{Hl} \frac{\partial H^l_Q}{\partial x^k} \tag{54}$$

The correction items newly introduced to classical Maxwell equations are highly non-linear as the local average rotation angles are defined by equations (23) and (35) in form.

The most important feature of the reformulated Maxwell equations is the electromagnetic field has limited "memory". Considering the lower index 0 corresponds to a reference time $t_0$, the reformed Maxwell equations show that its effects will still affect the electromagnetic field at time $t$. This time related "memory" effects will be named as "time-memory" here after.

As the lower index Q corresponds to the points on boundary, the reformed Maxwell equations show that its effects will still affect the electromagnetic field away from boundary. This boundary related "memory" effects will be named as "boundary-memory" here after.

The above two features make the reformed Maxwell equations be global equations rather than local equations.

For example of Maxwell equations with boundary memory condition, recently R Nibbi and S Polidora have shown an example[9].

## 4. Application in Superconductor

For superconductors, taking vacuum parameters $\varepsilon = \varepsilon_0$, $\mu = \mu_0$, we have:

$$\vec{D} = \varepsilon_0 \vec{E} = 0, \quad \vec{B} = \mu_0 \vec{H} = 0, \text{ within superconductor} \tag{55}$$

This shows us that for superconductors, the time memory effects can be omitted, and the boundary effects will take the main role.

Supposing the magnetic decaying rapidly along the boundary face in $j$ direction with a coherent length $\lambda_{Jos}^j$, for other case define the $\lambda_{Jos}^j \to \infty$, then the tensor

$$B_{Qj}^i = \frac{\partial B^i}{\partial B_Q^j}, \quad E_{Qj}^i = \frac{\partial E^i}{\partial E_Q^j} \tag{56}$$

will be completely determined by the material feature of superconductor and the boundary geometry.

The spatial differential will be approximated by:

$$\frac{\partial B_Q^i}{\partial x^j} \approx \frac{-1}{\lambda_{Jos}^j} B_Q^i \tag{57}$$

Hence, omitting higher infinitesimal, the related equations (51)-(54) can be rewritten as:

$$\sigma_B = (\cos\theta_Q)^{-1} R_{Bj}^i \frac{\partial B_Q^j}{\partial x^i} = -(\cos\theta_Q)^{-1} \frac{R_{Bj}^i B_Q^j}{\lambda_{Jos}^j} \tag{58}$$

$$J_E^i = (\cos\theta_E)^{-1} \cdot e_{ijk} R_{El}^j \frac{\partial E_Q^l}{\partial x^k} = (\cos\theta_E)^{-1} \frac{e_{ijk} R_{El}^j B_Q^l}{\lambda_{Jos}^k} \tag{59}$$

$$\sigma_D = (\cos\theta_D)^{-1} R_{Dj}^i \frac{\partial D_Q^j}{\partial x^i} = (\cos\theta_D)^{-1} \frac{R_{Dj}^i D_Q^j}{\lambda_{Jos}^i} \tag{60}$$

$$J_H^i = (\cos\theta_H)^{-1} \cdot e_{ijk} R_{Hl}^j \frac{\partial H_Q^l}{\partial x^k} = (\cos\theta_H)^{-1} \cdot e_{ijk} \frac{R_{Hl}^j H_Q^l}{\lambda_{Jos}^k} \tag{61}$$

For equation (61), omitting higher order infinitesimal, we will have:

$$\nabla \times \vec{J}_H = -(\cos\theta_H)^{-1} \nabla \theta_H \times \vec{J}_{QH} \tag{62}$$

where,

$$J_{QH}^i = e_{ijk} \frac{R_{Hl}^j}{\lambda_{Jos}^k} H_Q^l \tag{63}$$

Hence, we get:

$$(\nabla \times \vec{J}_H)^i = K_{QH}^{il} H_Q^l \tag{64}$$

As the $K_{QH}^{il}$ is determined by the material feature and geometry of superconductor, the equation (64) in fact is the general form of the London second equation.

Here, $\hat{R}_j^i$ is the orientation unit rotation tensor.

For superconductors

$$\nabla \cdot \vec{H} = (\cos\theta_H)^{-1} R_{Hj}^i \frac{\partial H_Q^j}{\partial x^i} \tag{58}$$

By equation (45),

$$\nabla \times \vec{J} = \nabla(\nabla \cdot \vec{H}) - \nabla^2 \vec{H} - \nabla \times \frac{\partial \vec{D}}{\partial t} \tag{59}$$

Omitting higher order infinitesimal, one gets:

$$\nabla \times \vec{J} = \nabla[(\cos\theta_H)^{-1} R_{Hj}^i \frac{\partial H_Q^j}{\partial x^i}] \tag{60}$$

Within a coherence length $\lambda_{jose}$ away from boundary, the fields decay into zero rapidly, hence we can write down:

$$\nabla \times \vec{J} = -\nabla[(\cos\theta_H)^{-1} R_{Hj}^i \hat{R}_i^l \frac{1}{\lambda_{Jose}^l} H_Q^j] \tag{61}$$

For simplest case, selecting the coordinator direction to make $\hat{R}_j^i = \delta_j^i$, we have:

$$(\cos\theta_H)^{-1} R_{Hj}^i \hat{R}_i^l \frac{1}{\lambda_{Jose}^l} H_Q^j = (\cos\theta_H)^{-1} R_{Hj}^i \frac{H_Q^j}{\lambda_{Jose}} \tag{62}$$

Although the local average rotation is small, but its change is rapidly, hence the above equation can be approximated as:

$$\nabla \times \vec{J} = -\nabla[(\cos\theta_H)^{-1}] \cdot \frac{1}{\lambda_{Jose}^n} H_Q^n \tag{63}$$

here, $n$ index represents boundary surface normal direction. As the current is along the boundary face, along the surface the $\nabla[(\cos\theta_H)^{-1}]$ only have normal component, hence, it can be represented as:

$$\nabla \times \vec{J} = -\alpha \vec{H} \tag{64}$$

where,

$$\alpha = \frac{\partial}{\partial n}[(\cos\theta_H)^{-1}] \cdot \frac{1}{\lambda_{Jose}^n} \tag{65}$$

This is the usual second London equation [10].
By equation (43), we have:

$$(\nabla \times \vec{E})^i = e_{ijk} \frac{\partial E^j}{\partial x^k} = (\cos\theta_E)^{-1} \cdot e_{ijk} R_{El}^j \frac{\partial E_Q^l}{\partial x^k} = -(\frac{\partial \vec{B}}{\partial t})^i = \frac{1}{\alpha} \frac{\partial}{\partial t}(\nabla \times \vec{J})^i \tag{66}$$

Hence, we get:

$$\vec{E} = \frac{1}{\alpha} \frac{\partial \vec{J}}{\partial t} \tag{67}$$

This is the first London equation [10].
It shows that the Josephson current is determined by boundary discontinuity features which depend on the boundary geometry and the material feature of superconductor.
For superconductors, to introduce Josephson current, the simplest way is to define:

$$\nabla \cdot \vec{B} = (\cos\theta_B)^{-1} R_{Bj}^i \frac{\partial B_Q^j}{\partial x^i} = \sigma_m \tag{68}$$

Then,

$$\nabla \times \vec{J} = \nabla[(\cos\theta_H)^{-1} R_{Hj}^i \frac{\partial H_Q^j}{\partial x^i}] = \frac{1}{\mu_0} \nabla[(\cos\theta_B)^{-1} R_{Bj}^i \frac{\partial B_{BQ}^j}{\partial x^i}] = \frac{1}{\mu_0} \nabla \sigma_m \tag{69}$$

Based on above research, the Josephson current is indeed produced by the local rotation of magnetic field near boundary (relative to outside magnetic field). This rotation corresponds to the curvature of magnetic line passing the boundary and magnitude change. Hence, it can be explained by the vortices of magnetic field near boundary. The equivalent radium of the local rotation of magnetic field can be used to represents the feature of rotation in concepts. But such a complicated theoretic interpretation cannot meet the demands of mathematic modeling which are strongly excited by the industrial development of superconductor-related devices.

The reformulated Maxwell equations show that if the magnetic field in superconductor is zero when outside magnetic field is $\vec{B}_c$, then when the outside magnetic field is zero the magnetic field in superconductor can be maintain as $\vec{B}_c$. This means magnetic field can be "iced" in the superconductor. Its reverse is also true, that means the magnetic field can be "pushed out" from superconductor.

By equation (68), $\sigma_m$ is determined by boundary condition. So it is reasonable to name it as boundary magnetic charge density. The reformulated Maxwell equations may fill up the gap between the industrial demands and the theoretic expression of magnetic field in superconductor.

**5. Conclusion**

The benefits of the reformulated Maxwell equation is the boundary and initial fields can be calculated by boundary condition and initial condition which in engineering design is usually pre-settled down. In most cases which related with industrial application, the initial and boundary condition can be measured, hence the reformulated Maxwell equation will be solvable. The more important point is that it bridges the gap between the classical Maxwell equation and the initial and boundary condition that may be logic un-consistent in that it causes un-solvability of the given problem.